\documentclass[12pt]{article}

\usepackage{geometry}
\usepackage{graphicx}
\usepackage{amsmath,amssymb,bm}
\usepackage{mathptmx}
\usepackage{color,nameref}
\usepackage[colorlinks,citecolor=blue,urlcolor=blue,filecolor=blue]{hyperref}
\usepackage{url}
\usepackage{natbib}
\usepackage{upgreek}
\usepackage{textgreek}
\usepackage{bm}
\newcommand{\nt}{\textbf{\fontfamily{phv}\selectfont\straighttheta}}
\newcommand{\nm}{\boldsymbol\mu}
\newcommand{\nS}{\boldsymbol\Sigma}
\newcommand{\EE}{\mathrm{E}}
\newcommand{\DD}{D}
\newcommand{\dd}{\mathrm{d}}
\newcommand{\ZZ}{\mathbf{Z}}
\newcommand{\CC}{\textnormal{C}}
\newcommand{\YY}{\textnormal{Y}}
\newcommand{\cov}{\textnormal{cov}}
\newcommand{\var}{\textnormal{var}}
\newcommand{\NS}{\gamma}
\long\def\comment#1{}

\begin{document}

\title{\Huge Spatial Statistics}

\author{Noel Cressie
\and Matthew T. Moores}

\date{\small{Centre for Environmental Informatics \\National Institute for Applied Statistics Research Australia (NIASRA) \\School of Mathematics \& Applied Statistics \\University of Wollongong, NSW 2522, Australia}\\ \vspace{15pt}
\today}

\maketitle

\begin{abstract}
Spatial statistics is an area of study devoted to the statistical analysis of data that have a spatial label associated with them. Geographers often refer to the ``location information'' associated with the ``attribute information,'' whose study defines a research area called ``spatial analysis.'' Many of the ways to manipulate spatial data are driven by algorithms with no uncertainty quantification associated with them. When a spatial analysis is statistical, that is, it incorporates uncertainty quantification, it falls in the research area called spatial statistics. The primary feature of spatial statistical models is that nearby attribute values are more statistically dependent than distant attribute values; this is a paraphrasing of what is sometimes called the First Law of Geography \citep{Tobler1970}.
\end{abstract}

\section{Introduction}\label{s:intro}
Spatial statistics provides a probabilistic framework for giving answers to those scientific questions  where  spatial-location information is present in the data, and that information is relevant to the questions being asked. The role of probability theory in (spatial) statistics is to model the uncertainty, both in the scientific theory behind the question, and in the (spatial) data coming from measurements of the (spatial) process that is a representation of the scientific theory.

In spatial statistics, uncertainty in the scientific theory is expressed probabilistically through a spatial stochastic process, which can be written most generally as: 
\begin{equation} 
\{Y(\mathbf{s}): \mathbf{s} \in \mathcal{D}\},                 \label{1aaa}
\end{equation}
where $Y(\mathbf{s})$ is the random  attribute value at location $\mathbf{s}$, and $\mathcal{D}$ is a subset of a d-dimensional space, here Euclidean space $\mathbb{R}^d$, that indexes all possible spatial locations of interest. Contained within $\mathcal{D}$ is a (possibly random) set $D$ that indexes those parts of $\mathcal{D}$ relevant to the scientific study. We shall see below that $D$ can have different set properties, depending upon whether the spatial process is a geostatistical process, a lattice process, or a point process. 

It is convenient to express the joint probability model defined by random $\{\YY (\mathbf{s}): \mathbf{s} \in \DD \}$ and random $\DD$ in the following shorthand, $[\YY, \DD]$, which we refer to as the spatial process model. Now,
\begin{equation}
[Y,\DD] = [Y\mid\DD][\DD],                         \label{2bbb}
\end{equation}
where for generic random quantities $A$ and $B$, their joint probability measure is denoted by $[A,\ B]$; the conditional probability measure of $A$ given $B$ is denoted by $[A \mid B]$; and the marginal probability measure of $B$ is denoted by $[B]$. In this review of spatial statistics, expression \eqref{2bbb} formalizes the general definition of a spatial statistical model given in Cressie (1993, Section 1.1).

The model \eqref{2bbb} covers the three principal spatial statistical areas according to three different assumptions about $[\DD]$, which leads to three different types of spatial stochastic process, $[Y \mid \DD$]; these are  described further in the next section, titled ``\nameref{s:model}.'' Spatial statistics has, in the past, classified its methodology according to the types of spatial data, denoted here as $\ZZ$ \citetext{e.g., \citealp{Ripley1981,upton1985spatial}; and \citealp{Cressie1993}}, rather than the types of spatial processes $Y$ that underly the spatial data.  

In this review, we classify our spatial-statistical modeling choices according to the \textit{process model} \eqref{2bbb}. Then the \textit{data model}, namely the distribution of the data $\ZZ$ given both $Y$ and $D$ in \eqref{2bbb}, is the straightforward conditional-probability measure,
\begin{equation}
[\mathbf{Z} \mid Y, \DD] .                 \label{3ccc}
\end{equation}
For example, the spatial data $\ZZ$ could be the vector $\left(Z(\mathbf{s}_1), ..., Z(\mathbf{s}_n)\right)'$, of imperfect measurements of $Y$ taken at given spatial locations $\{ \mathbf{s}_1, \dots, \mathbf{s}_n \} \subset \DD$, where the data are assumed to be \textit{conditionally independent}. That is, the data model is
\begin{equation}
[\mathbf{Z}\mid Y, \DD] = \prod_{i=1}^{n} [Z(\mathbf{s}_i) \mid Y, \DD].              \label{4ddd}
\end{equation}
Notice that while \eqref{4ddd} is based on conditional independence, the {\em marginal} distribution, $[\ZZ \mid \DD]$, does \textit{not} exhibit independence: The spatial-statistical dependence in $\ZZ$, articulated in the First Law of Geography that was discussed in the abstract, is inherited from  $[Y \mid \DD]$ and \eqref{4ddd} as follows: 
\begin{equation*}
[\mathbf{Z} \mid \DD] = \int[\mathbf{Z} \mid Y, \DD][Y \mid \DD] \ \dd Y.   \\    
\end{equation*}

Another example is where the randomness is in $\DD$ but not in $Y$. If $\DD$ is a point process (a special case of a random set), then the data $\mathbf{Z} = \{N,\mathbf{s}_1,\dots, \mathbf{s}_N\}$, where $N$ is the random number of points in the now-bounded region $\mathcal{D}$, and $D$ = \{$\mathbf{s}_1,\dots, \mathbf{s}_N$\} are the random locations of the points. If there are measurements (sometimes called ``marks'') \{$Z(\mathbf{s}_1),...,Z(\mathbf{s}_N)$\} associated with the random points in $D$, these should be included within \textbf{Z}. That is,
\begin{equation}
 \mathbf{Z} = \left\{N, \left(\mathbf{s}_1, Z (\mathbf{s}_1)\right), \dots,  \left(\mathbf{s}_N, {Z} (\mathbf{s}_N)\right)\right\}.     \label{eq:markedPoint}
\end{equation}

This description of spatial statistics given by \eqref{2bbb} and \eqref{3ccc} captures the (known) uncertainty in the scientific problem being addressed, namely \textit{scientific uncertainty} through the spatial process model \eqref{2bbb} and \textit{measurement uncertainty} through the data model \eqref{3ccc}. Together, \eqref{2bbb} and \eqref{3ccc} define a \textit{hierarchical statistical model}, here for spatial data, although this hierarchical formulation through the conditional probability distributions,  $[\mathbf{Z} \mid Y, \DD], [Y \mid \DD],$ and $[\DD]$ for general $Y$ and $D$, is appropriate throughout all of applied statistics. 

It is implicit in (2) and (3) that any parameters $\nt$ associated with the process model and the data model are known. We now discuss how to handle parameter uncertainty in the hierarchical statistical model. A Bayesian would put a probability distribution on $\nt$ : Let $[\nt]$ denote the parameter model (or prior) that captures parameter uncertainty. Then, using obvious notation, all the uncertainty in the problem is  expressed through the joint probability measure,
\begin{eqnarray}
[\mathbf{Z}, Y, \DD, \nt] &=& [\mathbf{Z}, Y, \DD \mid \nt] [\nt] \label{eq:j2} \\
&=& [\mathbf{Z} \mid Y,\DD,{\nt}][Y \mid \DD, \nt ][\DD \mid \nt] [\nt].  \label{eq:joint}   
\end{eqnarray}
A {\em Bayesian hierarchical model} uses the decomposition \eqref{eq:joint}, but there is also an {\em empirical hierarchical model} that substitutes a point estimate $\hat\nt$ of $\nt$ into the first factor on the left-hand side of \eqref{eq:j2}, resulting in its being written as,
\begin{eqnarray}
[\mathbf{Z}, Y, \DD \mid \hat\nt ] = [{\mathbf{Z} \mid Y, D,\hat\nt ] [Y \mid  \DD, \hat\nt ] [D \mid \hat\nt }].   \label{6e}
\end{eqnarray}

Finding efficient estimators of $\nt$ from the spatial data \textbf{Z} is an important problem in spatial statistics, but in this review we emphasize the problem of spatial prediction of $Y$. In what follows, we shall  assume that the parameters are either known or have been estimated. Hence, for convenience, we can drop $\hat\nt$ in \eqref{6e} and observe that the uncertainty in the problem is expressed through the joint probability measure,
\begin{eqnarray}
[\mathbf{Z}, Y, \DD] = [\mathbf{Z} \mid Y,\DD] [Y \mid \DD] [\DD],            \label{7a} 
\end{eqnarray}
and Bayes' Rule can be used to infer the unknowns $Y$ and $\DD$ through the \textit{predictive distribution}: 
\begin{eqnarray}
[Y,\DD \mid \mathbf{Z}] = \frac{[\mathbf{Z} \mid Y,\DD] [Y \mid \DD] [\DD]}{[\mathbf{Z}]} ,        \label{8a}
\end{eqnarray}
where $[\mathbf{Z}]$ is the normalization constant that ensures that the right-hand side of \eqref{8a} integrates or sums to 1.      
If the spatial index set $D$ \textit{is fixed and known} then we can drop $D$ from \eqref{8a}, and Bayes' Rule simplifies to:
\begin{eqnarray}
[Y \mid \mathbf{Z}] = \frac{[\mathbf{Z} \mid Y] [Y ]}{[\mathbf{Z}]},        \label{9a}
\end{eqnarray}
which is the \textit{predictive distribution} of $Y$ (when $D$ is fixed and known). It is this expression that is often used in spatial statistics for prediction. For example, the well known \textit{simple kriging predictor} can easily be identified as the predictive mean of \eqref{9a} under Gaussian distributional assumptions for both \eqref{2bbb} and \eqref{3ccc} (Cressie and Wikle, 2011, pp. 139-141).

Our review of spatial statistics starts with a presentation in the next section, ``\nameref{s:model},'' of a number of commonly used spatial process models, which includes multivariate  models. Following that, the section, ``\nameref{s:discrete},'' turns attention to discretization of  $\mathcal{D} \subset \mathbb{R}^d$, which is an extremely important consideration when actually computing the predictive distribution \eqref{8a} or \eqref{9a}. The extension of spatial process models to spatio-temporal process models is discussed in the section, ``\nameref{s:spacetime}.'' Finally, in the ``\nameref{s:trends}'' section, we briefly discuss important recent research topics in spatial statistics, but due to a lack of space we are unable to present them in full. 
It will be interesting to see ten years from now, how these topics have evolved.

\section{Spatial Process Models}\label{s:model}

In this section, we set out various ways that the probability distributions $[Y \mid \DD]$, $[Y]$, and $[\DD]$, given in Bayes' Rule \eqref{8a}, can be represented in the spatial context. These are not to be confused with $[\mathbf{Z} \mid \DD]$ and $[\mathbf{Z}]$, the probability distributions of the spatial data. In many parts of the spatial-statistics literature, this confusion is noticeable when researchers build models directly for $[\mathbf{Z} \mid \DD]$. Taking a hierarchical approach, we capture knowledge of the scientific process starting with the statistical models, $[Y \mid \DD]$ and $[\DD]$, and then we model the measurement errors and missing data through $[\mathbf{Z} \mid Y, \DD]$. Finally, Bayes' Rule \eqref{8a} allows inference on the unknowns $Y$ and $\DD$ through the predictive distribution, $[Y,\DD \mid \mathbf{Z}]$.
   
We present three types of spatial process models, where their distinction is made according to the index set $\DD$ of all spatial locations at which the process $Y$ is defined. For a {\em geostatistical process}, $\DD = \DD^G$, which is a known set over which the locations vary continuously and whose area (or volume) is \textgreater \ 0. For a {\em lattice process}, $\DD  = \DD^L$, which is a known set whose locations vary discretely and the number of locations is countable; note that the area of $\DD^L$ is equal to zero. For a {\em point process}, $\DD = \DD^P$, which is a random set made up of random points in $\mathbb{R}^d$.

\subsection{Geostatistical Processes}\label{s:geostat} 
In this section, we assume that the spatial locations $\DD$ are given by $\DD^G$,  where $\DD^G$ is known. Hence $\DD$ can be dropped from any of the probability distributions in \eqref{8a}, resulting in \eqref{9a}. This allows us to concentrate on  $Y$ and, to feature the spatial index, we write $Y$ equivalently as $\{Y(\mathbf{s}):\mathbf{s}  \in D^G\}$. A property of geostatistical processes is that $D^G$ has positive area and hence is uncountable.

\begin{figure}
    \centering
    \includegraphics[width=\textwidth]{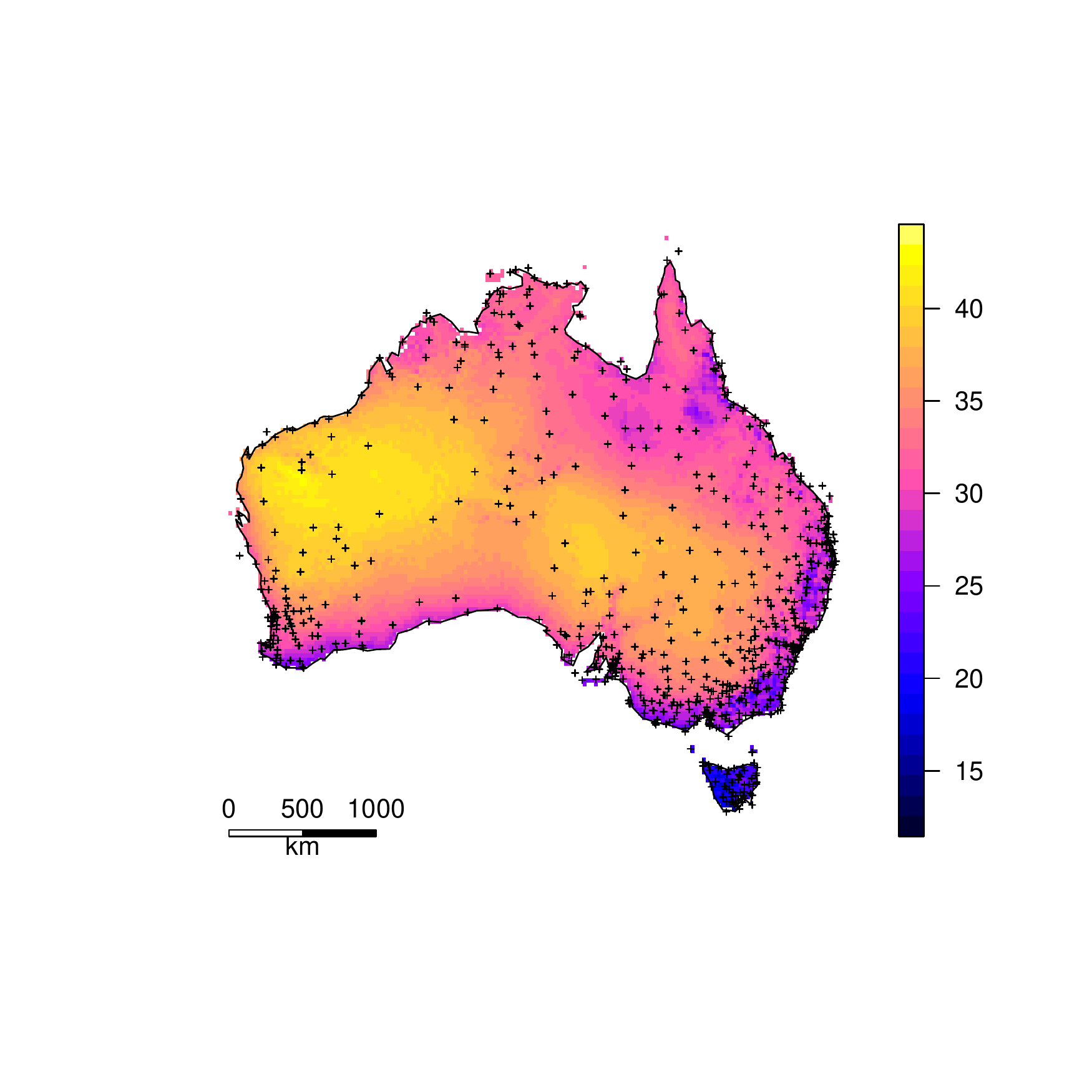}
    \caption{Map of a kriging predictor of Australian temperature in January 2009, superimposed on spatial locations of data.}\label{f:Kriging}
\end{figure}

Traditionally, a geostatistical process has been specified up to second moments. Starting with the most general specification, we have 
\begin{eqnarray}
\mu_Y{(\mathbf{s})} &\equiv&  \EE\left(Y(\mathbf{s})\right); \;\;\;  \mathbf{s} \in \DD^G  \label{b21a}  \\   
\CC_Y  {(\mathbf{s}, \mathbf{u})} &\equiv& \cov {(Y(\mathbf{s}), Y(\mathbf{u})); \;\;\; \mathbf{s}, \mathbf{u} \in \DD^G}.      \label{b21b}
\end{eqnarray}

From \eqref{b21a} and \eqref{b21b}, an optimal spatial linear predictor $\hat{Y}(\mathbf{s}_0)$ of ${Y}(\mathbf{s}_0)$, can be obtained that depends on spatial data $\mathbf{Z} \equiv \ (Z(\mathbf{s}_1),...,Z(\mathbf{s}_n))'$. This is an $n$-dimensional vector indexed by the data's $n$ known spatial locations, $\DD ^{G*} \equiv \ \{\mathbf{s}_1, \dots, \mathbf{s}_n \}\ \subset \DD^G$. In practice, estimation of the parameters $\nt$ that specify completely \eqref{b21a} and \eqref{b21b} can be problematic due to the lack of replicated data, so \cite{Matheron1963} made  stationarity assumptions that together are now known as \textit{intrinsic stationarity}. That is, for all $\mathbf{s,u} \in {\DD^G}$, assume
\begin{eqnarray}  
E(Y(\mathbf{s})) &=& \mu_Y^o        \label{c21a}             \\   
\var\left(Y(\mathbf{s}) - Y(\mathbf{u}) \right) &=& 2\NS^{o}_Y (\mathbf{s-u})  ,      \label{c21c}     
\end{eqnarray}
where \eqref{c21c} is equal to $ \CC_Y (\mathbf{s,s})+\CC_Y (\mathbf{u,u}) - 2 \CC_Y (\mathbf{s,u)}$. The quantity $2\NS^{o}_Y(\cdot)$ is called the \textit{variogram}, and $\NS^{o}_Y(\cdot)$ is called the \textit{semivariogram} (or occasionally the semivariance).

If the assumption in \eqref{c21c} were replaced by
\begin{eqnarray}
\cov (Y(\mathbf{s}), Y(\mathbf{u}) = \CC_Y^o (\mathbf{s-u}) ,\ \text{for all}  \ {\mathbf{s,u} \in \DD^G},           \label{a21c}           
\end{eqnarray}
\noindent then \eqref{a21c} and \eqref{c21a}  together are known as {\em second-order stationarity}. Matheron chose \eqref{c21c} because he could derive optimal-spatial-linear-prediction (i.e., kriging) equations of ${Y}(\mathbf{s}_0)$ without having to know or estimate $\mu_Y^o$. Here,  ``optimal'' is in reference to a spatial linear predictor $\hat{Y}(\mathbf{s}_0)$  that minimizes the mean-squared prediction error (MSPE),
\begin{equation}
\EE \left[\left(\hat{Y}(\mathbf{s}_0) - {Y}(\mathbf{s}_0)\right)^2\right] , \text{ for any } \mathbf{s}_0 \in D^G ,  \label{d21d}
\end{equation}
where $\hat{Y}(\mathbf{s}_0) \equiv \sum_{i=1}^n \lambda_i Z(\mathbf{s}_i)$. The minimization in \eqref{d21d} is with respect to the coefficients $\{\lambda_i\ : i = 1,...,n\}$ subject to the unbiasedness constraint, $\EE \left(\hat{Y}(\mathbf{s}_0)\right) = \EE \left({Y}(\mathbf{s}_0)\right)$, or equivalently subject to the constraint $\sum_{i=1}^n \lambda_i = 1$ on $\{\lambda_i\}$. With optimally chosen $\{ \lambda_i \}$, $\hat{Y}(\mathbf{s}_0)$ is known as the {\em kriging predictor}. 
Matheron called this approach to spatial prediction {\em ordinary kriging}, although it is known in other fields as BLUP (Best Linear Unbiased Prediction); \citet{Cressie1990} gave the history of kriging and showed that it could also be referred to  descriptively as \textit{spatial BLUP}. 

\begin{figure}
    \centering
    \includegraphics[width=\textwidth]{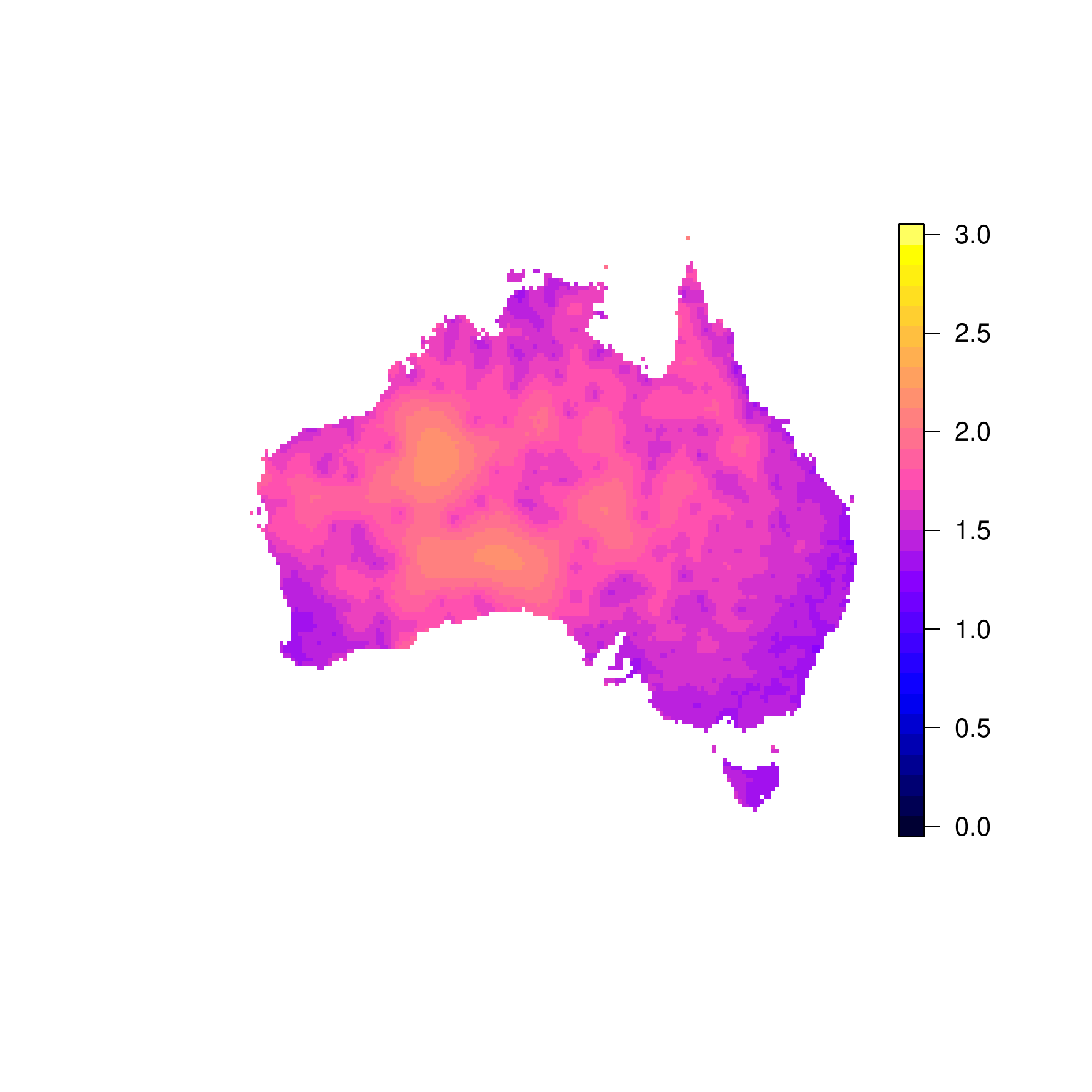}
    \caption{Map of the kriging standard error \eqref{eq:krigSE} for the kriging predictor shown in Figure~\ref{f:Kriging}.}
		\label{f:Map}
\end{figure}

The constant-mean assumption \eqref{c21a} can be generalized to $E (Y(\mathbf{s})) \equiv \mathbf{x}(\mathbf{s})'\boldsymbol\beta$, for $\mathbf{s} \in \DD^G$, which is a linear regression where the regression coefficients $\boldsymbol\beta$ are unknown and the covariate vector $\mathbf{x}(\mathbf{s})$ includes the entry 1. Under this assumption on $\EE \left({Y}(\mathbf{s})\right)$, ordinary kriging is generalized to {\em universal kriging}, also notated as $\hat{Y}(\mathbf{s}_0)$.
Figure~\ref{f:Kriging} shows the universal-kriging predictor of Australian temperature in the month of January 2009, mapped over the whole continent $\DD^G$, where the spatial locations $D^{G*} = \{ \mathbf{s}_1, \dots, \mathbf{s}_n \}$ of weather stations that supplied the data $\mathbf{Z}$ are superimposed. Formulas for $\hat{Y}(\mathbf{s}_0)$ can be found in, for example, \citet[][Section 3.4]{chiles2009geostatistics}.

The optimized MSPE \eqref{d21d} is called the {\em kriging variance}, and its square root is called the {\em kriging standard error}:
\begin{equation} 
\sigma_k (\mathbf{s}_0) \equiv \left(\EE \left(\hat{Y}(\mathbf{s}_0) - {Y}(\mathbf{s}_0)\right)^2\right)^{1/2}, \ \text{for any} \ \mathbf{s}_0 \in \DD^G. \label{eq:krigSE}
\end{equation}
Figure~\ref{f:Map} shows a map over $\DD^G$ of the kriging standard error associated with the kriging predictor mapped in Figure~\ref{f:Kriging}. It can be shown that a smaller $\sigma_k (\mathbf{s}_0)$ corresponds to a higher density of weather stations near $\mathbf{s}_0$. 
While ordinary and universal kriging produce an optimal linear predictor, there is an even better predictor, the {\em best optimal predictor} ({\em BOP}), which is the best
of all the best predictors obtained under extra constraints (e.g., linearity). From Bayes' Rule \eqref{8a}, the predictor that minimizes the MSPE \eqref{a21c} without any constraints is $Y^*(\mathbf{s}_0) \equiv \EE\left(Y(\mathbf{s}_0) \mid \mathbf{Z} \right)$, which is the mean of the predictive distribution. Notice that the BOP, $Y^*(\mathbf{s}_0)$, is unbiased, namely $\EE(Y^*(\mathbf{s}_0)) = \EE(Y(\mathbf{s}_0))$, without having to constrain it to be so.

\subsection{Lattice Processes}\label{s:lattice}

In this section, we assume that the spatial locations $\DD$ are given by $\DD^L$, a known countable subset of $\mathbb{R}^d$. This  usually represents a collection of grid nodes, pixels, or small areas and the spatial locations associated with them; we write the countable set of all such locations as ${\DD}^L \equiv \left\{ \mathbf{s}_1, \mathbf{s}_2, \dots \right\}$. Each $\mathbf{s}_i$ has a set of neighbors, $\mathcal{N}(\mathbf{s}_i) \subset \DD^L \setminus \mathbf{s}_i$, associated with it, and whose locations are spatially proximate (and note that a location is not considered to be its own neighbor). Spatial-statistical dependence between locations in lattice processes is  defined in terms of these neighborhood relations. 

Typically, the neighbors are represented by a spatial-dependence matrix $\mathbf{W}$ with entries $w_{i,j}$ nonzero if $\mathbf{s}_j \in \mathcal{N}(\mathbf{s}_i)$, and hence the diagonal entries of $\mathbf{W}$ are all zero. The non-diagonal entries of $\mathbf{W}$ might be, for example, inversely proportional to the distance, $\| \mathbf{s}_i - \mathbf{s}_j \|$, or they might involve some other way of moderating dependence based on spatial proximity. For example, they might be assigned the value 1 if a neighborhood relation exists and 0 otherwise. In this case, $\mathbf{W}$ is called an adjacency matrix, and it is symmetric if $\mathbf{s}_j \in \mathcal{N}(\mathbf{s}_i)$ whenever  $\mathbf{s}_i \in \mathcal{N}(\mathbf{s}_j)$ and {\em vice versa}. 

Consider a lattice process in $\mathbb{R}^2$ defined on the finite grid ${\DD}^L = \{ (x, y) : x,y = 1,\dots,5 \}$. The first-order neighbors of the grid node $(x,y)$ in the interior of the lattice are the four adjacent nodes, $\mathcal{N}(x,y) = \{ (x-1, y), (x, y-1), (x+1, y), (x, y+1) \}$, shown as:
$$
\arraycolsep=4pt\def\arraystretch{1}\begin{array}{ c c c c c }
\circ & \circ & \circ & \circ & \circ \\
\circ & \circ & {\color{blue} \bullet} & \circ & \circ \\
\circ & {\color{blue} \bullet} & {\color{red} \boldsymbol\times} & {\color{blue} \bullet} & \circ \\
\circ & \circ & {\color{blue} \bullet} & \circ & \circ \\
\circ & \circ & \circ & \circ & \circ
\end{array}
$$ 
where grid node $\mathbf{s}_i$ is represented by ${\color{red} \boldsymbol\times}$, and its first-order neighbors are represented by ${\color{blue} \bullet}$. Nodes ${\color{red} \boldsymbol\times}$ situated on the boundary of the grid will have less than four neighbours.

The most common type of lattice process is the {\em Markov random field} (MRF), which has a conditional-probability property in the spatial domain $\mathbb{R}^d$ that is a generalization of the temporal Markov property found in section, ``\nameref{s:spacetime}.'' A lattice process $\{ Y(\mathbf{s}) : \mathbf{s} \in \DD^L \}$ is a MRF if, for all $\mathbf{s}_i \in \DD^L$, its conditional probabilities satisfy 
\begin{equation}
\left[ Y(\mathbf{s}_i) \mid \mathbf{Y}({\DD}^L \setminus \mathbf{s}_i) \right] = \left[ Y(\mathbf{s}_i) \mid \mathbf{Y}(\mathcal{N}(\mathbf{s}_i)) \right], \label{17mtm}
\end{equation} 
where $\mathbf{Y}(A) \equiv \{ Y(\mathbf{s}_j) : \mathbf{s}_j \in A \}$. The MRF is defined in terms of these conditional probabilities \eqref{17mtm}, which represent statistical dependencies between neighbouring nodes that are captured differently from those given by the variogram or the covariance function. Specifically, 
\begin{equation}
\label{eq:lit_mrf}
\left[ Y(\mathbf{s}_i) \mid \mathbf{Y}({\DD}^L \setminus \mathbf{s}_i) \right] \;=\; \frac{\exp\left\{-f (Y(\mathbf{s}_i), \mathbf{Y}(\mathcal{N}(\mathbf{s}_i))) \right\}}{\CC},
\end{equation}
where $\CC$ is a normalizing constant that ensures the right-hand side of \eqref{eq:lit_mrf} integrates (or sums) to 1. Equation \eqref{eq:lit_mrf} is also known as a Gibbs random field in statistical mechanics since, under regularity conditions, the Hammersley-Clifford Theorem relates the joint probability distribution to the Gibbs measure \citep{Besag1974}. The function $f(Y(\mathbf{s}_i), \mathbf{Y}(\mathcal{N}(\mathbf{s}_i)))$ is referred to as the potential energy, since it quantifies the strength of interactions between neighbors. A wide variety of MRF models can be defined by choosing different forms of the potential-energy function \citep[][Section 3.2] {Winkler2003}. Note that care needs to be taken to ensure that specification of the model through all the conditional probability distributions, $\left\{ \left[ Y(\mathbf{s}_i) \mid \mathbf{Y}(\mathcal{N}(\mathbf{s}_i)) \right] : \mathbf{s}_i \in D^L \right\}$, results in a valid joint probability distribution, $\left[ \left\{ Y(\mathbf{s}_i) : \mathbf{s}_i \in D^L \right\} \right]$ \citep{kaiser2000construction}.

Revisiting the previous simple example of a first-order neighborhood structure on a regular lattice in $\mathbb{R}^2$, notice that grid nodes situated diagonally across from each other are conditionally independent. Hence, ${\DD}^L$ can be partitioned into two sub-lattices ${\DD}^L_1$ and ${\DD}^L_2$, such that the values at the nodes in ${\DD}^L_1$ are independent given the values at the nodes in ${\DD}^L_2$ and vice versa \citep[Section 8.1]{Besag1974,Winkler2003}:
$$
\arraycolsep=4pt\def\arraystretch{1}\begin{array}{ c c c c c }
\bullet & \circ & \bullet & \circ & \bullet \\
\circ & \bullet & \circ & \bullet & \circ \\
\bullet & \circ & \bullet & \circ & \bullet \\
\circ & \bullet & \circ & \bullet & \circ \\
\bullet & \circ & \bullet & \circ & \bullet \\
\end{array}
$$
This forms a checkerboard  pattern where $\left\{Y(\mathbf{s}):\mathbf{s} \in D^L_1 \right\}$ at nodes $D^L_1$ represented by $\bullet$ are mutually independent, given the values $\left\{Y(\mathbf{u}):\mathbf{u} \in D^L_2\right\}$ at nodes $D^L_2$ represented by $\circ$. 

\citet{Besag1974} introduced the conditional autoregressive (CAR) model, which is a Gaussian MRF that is defined in terms of its conditional means and variances. We refer the reader to \citet{LeSagePace2009} for discussion of a different lattice-process model, known as the simultaneous autoregressive (SAR) model, and a comparison of it with the CAR model. We define the CAR model as follows: For $\mathbf{s}_i \in D^L$, $Y(\mathbf{s}_i)$ is conditionally Gaussian defined by its first and second moments,
\begin{eqnarray}
\EE (Y(\mathbf{s}_i) \mid \mathbf{Y}(\mathcal{N}(\mathbf{s}_i))) &=& \sum_{\mathbf{s}_j \in \mathcal{N}(\mathbf{s}_i)} c_{i,j} Y(\mathbf{s}_j) \label{mrf_mean} \\
\var (Y(\mathbf{s}_i) \mid \mathbf{Y}(\mathcal{N}(\mathbf{s}_i))) &=& \tau^2_i ,
\end{eqnarray}
where $c_{i,j}$ are spatial autoregressive coefficients such that the diagonal elements $c_{1,1} = \dots = c_{n,n} = 0$, and $\{ \tau^2_i \}$ are the scale parameters for the locations $\{\mathbf{s}_i\}$, respectively. Under an important regularity condition (see below), this specification results in a joint probability distribution that is multivariate Gaussian. That is,
\begin{eqnarray}\label{eqMVN}
\mathbf{Y} \sim Gau (\mathbf{0}, (\mathbf{I} - \mathbf{C})^{-1} \mathbf{M}),
\end{eqnarray}
where Gau $(\nm,\nS)$ denotes a Gaussian distribution with mean vector $\nm$ and covariance matrix ${\nS}$;  the matrix $\mathbf{M} \equiv$ diag $(\tau^2_1, \dots, \tau^2_n)$ is diagonal; and the regularity condition referred to above is that the coefficients $\mathbf{C} \equiv \{ c_{i,j} \}$   in \eqref{mrf_mean} have to result in $\mathbf{M}^{-1} (\mathbf{I} - \mathbf{C})$ being a symmetric and positive-definite matrix. With a first-order neighborhood structure, such as shown in the simple example above in $\mathbb{R}^2$, the precision matrix is block-diagonal, which makes it possible to sample efficiently from this Gaussian MRF using sparse-matrix methods \citep[][Section 2.4]{Rue2005}.

The data vector for lattice processes is $\mathbf{Z} \equiv \left( Z(\mathbf{s}_1), \dots, Z(\mathbf{s}_n) \right)'$, where $D^{L*}\equiv \{ \mathbf{s}_1, \dots, \mathbf{s}_n \} \subset D^L$. As for the previous subsection, the data model is $[ \mathbf{Z} \mid \{ Y(\mathbf{s}) : \mathbf{s} \in D^L \} ]$ which, to emphasize dependance on parameters $\nt$, we reactivate earlier notation and write it as $[\mathbf{Z} \mid Y, \nt]$. Now, if we write the lattice-process model as $[ \{ Y(\mathbf{s}) : \mathbf{s} \in D^L \} \mid \nt ] \equiv [ Y \mid \nt ]$, then estimation of $\nt$ follows by maximizing the likelihood, $\mathcal{L}(\nt) \equiv \int \ [\mathbf{Z} \mid Y, \nt] \ [ Y \mid \nt ] \ \dd Y$. 

Regarding  spatial prediction, $Y^*(\mathbf{s}_0) \equiv E (Y(\mathbf{s}_0) \mid \mathbf{Z}, \nt)$ is the best optimal predictor of $Y(\mathbf{s}_0)$, for $\mathbf{s}_0 \in D^L_\cdot$ and known $\nt$   \citep[e.g., ][]{besag1991bayesian}. Note that $\mathbf{s}_0$ may not belong to $D^{L*}$, and hence $Y^*(\mathbf{s}_0)$ is a predictor of $Y (\mathbf{s}_0)$ even when there is no datum observed at the node $\mathbf{s}_0.$ Inference on unobserved parts of the process $Y$ is just as important for lattice processes as it is for geostatistical processes.

\subsection{Spatial Point Processes and Random Sets}\label{s:point}
A spatial point process is a countable collection of {\em random locations} $\DD \equiv \DD^P \subset \mathcal{D}$. Closely related to this random set of points is the counting process that we shall call $\{ N(A) : A \subset \mathcal{D} \}$, where recall that $\mathcal{D}$ indexes all possible locations of interest, and now we assume it is bounded. For example, if $A$ is a given subset of $\mathcal{D}$, and two of the random points $\left\{\mathbf{s}_i\right\}$ are contained in $A$, then $N(A) = 2$. Since $D^P =\left\{\mathbf{s}_i\right\}$ is random and $A$ is fixed, $N(A)$ is a random variable defined on the non-negative integers.

Clearly, the joint distributions $[N(A_1),\dots,N(A_m)]$, for any subsets $\left\{A_j :j = 1,\dots,m\right\}$ contained in $\mathcal{D}$ (possibly overlapping) and for any $m = 0,1,2,\dots,$ are well defined. Spatial dependence can be seen through the spatial proximity between the $\left\{A_j\right\}$. Consider just two fixed subsets, $A_1$ and $A_2$ (i.e., $m=2$) and, to avoid ambiguity caused by potentially sharing points, let $A_1 \cap  A_2$ be empty. Then no spatial dependence is exhibited if, for any disjoint $A_1$ and $A_2$, there is statistical independence; that is,
\begin{equation}
\left[N(A_1), N(A_2)\right] = \left[N(A_1)\right]\left[N(A_2)\right] .         \label{a23a}
\end{equation}
The basic point process known as the \textit{Poisson point process} has the independence property \eqref{a23a}, and its associated counting process satisfies
\begin{equation}
\left[N(A)\right] = \exp\{-\lambda(A)\} \frac{\lambda(A)^{N(A)}}{N(A)!} ; \;\;\; \ A \subset \mathcal{D} ,    \label{b23b}
\end{equation}
where $\lambda (A) \equiv \int_A \lambda(\mathbf{s}) d\mathbf{s}$. In \eqref{b23b}, $\lambda (\cdot)$ is a given intensity function defined according to:
\begin{equation}
\lambda(\mathbf{s}) \equiv \lim\limits_{ | \delta \mathbf{s} | \to 0} \frac{\EE\left( Y(\delta \mathbf{s})\right)}{| \delta \mathbf{s} |} ,
\end{equation}
where $\delta\mathbf{s}$ is a small set centered at $\mathbf{s} \in \mathcal{D}$, and whose volume is $| \delta \mathbf{s} |$.

\begin{figure}
    \centering
    \includegraphics[width=0.55\textwidth]{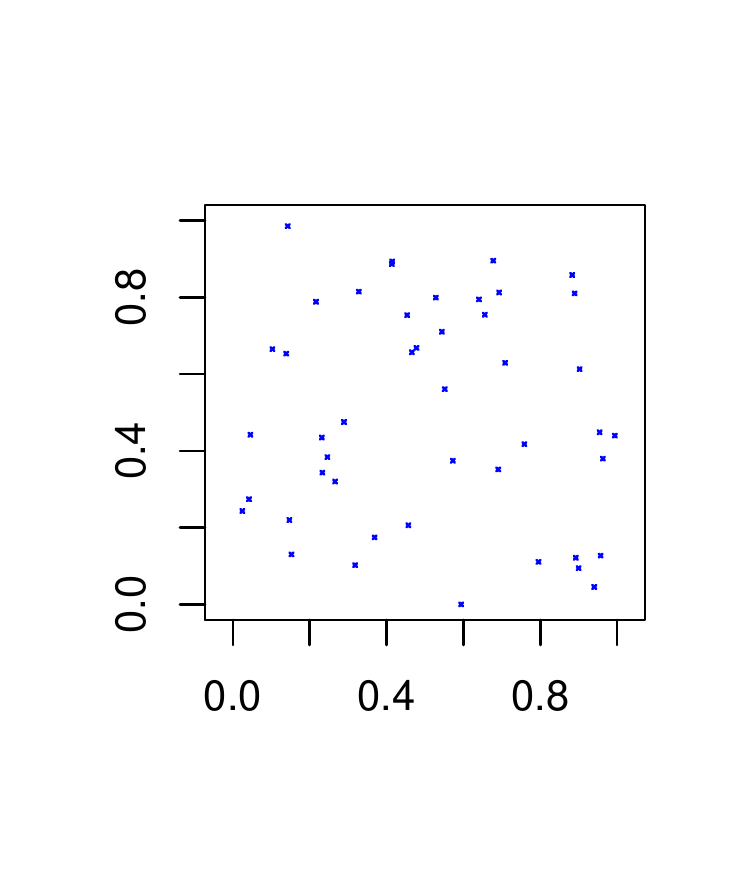}
    \caption{A realization on the unit square $\mathcal{D} = [0,1] \times [0,1]$, of the homogenous Poisson point process \eqref{b23b} with parameter $\lambda=50$; for this realization, $N(\mathcal{D}) = 46$.}\label{f:a32a}
\end{figure}

In \eqref{b23b}, the special case of $\lambda\mathbf({s})\equiv \lambda$, for all $\mathbf{s} \in \mathcal{D}$, results in a \textit{homogeneous Poisson point process}, and a simulation of it is shown in Figure \ref{f:a32a}. The simulation was obtained using an equivalent probabilistic representation for which the count random variable $N(\mathcal{D})$, for $\mathcal{D} = [0,1] \times [0,1]$,  was simulated according to \eqref{b23b}. Then, conditional on $N(\mathcal{D}), \{\mathbf{s}_1$, \dots, $\mathbf{s}_{N(\mathcal{D})}\}$ was simulated independently and identically according to the uniform distribution, 
\begin{equation}
[\mathbf{u}] = {\begin{cases} \frac{1}{ \lambda(\mathcal{D}) } \  ; \;\;\; {\mathbf{u} \in \mathcal{D}} \\  0  \ ; \;\;\; \text{elsewhere}. \end{cases}}
\end{equation}

This representation explains why the homogenous Poisson point process is commonly referred to as a {\em Completely Spatially Random} (CSR) process, and why it is used as a baseline for testing for the absence of spatial dependance in a point process. That is, before a spatial model is fitted to a point pattern, a test of the null hypothesis that the point pattern originates from a CSR process, is often carried out. Rejection of CSR then justifies the fitting of spatially dependent point processes to the data \citep[e.g., ][]{Ripley1981,Diggle2014}.

Much of the early research in point processes was devoted to establishing test statistics that were sensitive to various types of departures from the CSR process \citep[e.g., ][Section 8.2]{Cressie1993}. This was followed by researchers' defining and then estimating spatial-dependence measures such as the second-order-intensity function and the K-function  \citep[e.g.,][Chapter 8]{Ripley1981}, where inference was often in terms of method-of-moments estimation of these functions. More efficient likelihood-based inference came later; \citet{Baddeley2015} give a comprehensive review of these methodologies for point processes. From a modeling perspective, particular attention has been paid to the log Gaussian Cox point processes; here, $\lambda(\cdot)$ in \eqref{b23b} is random, such that $\left\{\text{log}  (\lambda(\mathbf{s})):\mathbf{s} \in \mathcal{D}\right\}$ is a Gaussian process \citep[e.g.,][]{Moller2003}. This model leads naturally to hierarchical Bayesian inference for $\lambda (\cdot)$ and its parameters \citep[e.g., ][]{Gelfand2018}. 

If an attribute process, $\{Y (\mathbf{s}_i): \mathbf{s}_i \in D^P \}$, is included with the spatial point process $D^P$, one obtains a so-called \textit{marked point process} \citep[e.g.,][Section 8.7]{Cressie1993}. For example, the study of a natural forest where both the locations $\{ \mathbf{s}_i \}$ and the sizes of the trees $\{ Y(\mathbf{s}_i) \}$ are modeled together probabilistically, results in a marked point process where the ``mark'' process is a spatial process $\{ Y(\mathbf{s}_i): \mathbf{s}_i \in \DD^P\}$ of tree size. Now, Bayes' Rule given by \eqref{8a}, where both $Y$ and $\DD$ $(= D^P)$ are random, should be used to make inference on $Y$ and $D^P$ through the predictive distribution $[Y, D^P \mid \mathbf{Z}]$. Here, $\mathbf{Z}$ consists of the number of trees, the trees' locations, and their size measurements, as denoted in \eqref{eq:markedPoint}. After marginalization, we can obtain $[\DD^P \mid \mathbf{Z}]$, the predictive distribution of the spatial point process $D^P$.

A spatial point process is a special case of a random set, which  is a random quantity in Euclidean space that was defined rigorously by \citet{Matheron1975}. Some geological processes are more naturally modeled as set-valued phenomena (e.g., the facies of a mineralization), however inference for random-set processes has lagged behind those for spatial point processes. It is difficult to define a likelihood based on set-valued data, which has held back statistically efficient inferences; nevertheless, basic method-of-moment estimators are often available. The most well known random set that allows statistical inference from set-valued data is the Boolean Model \citep[e.g.,][Section 4.4]{Cressie2011}.

\subsection{Multivariate Spatial Processes}\label{s:multivariate}

The previous subsections have presented single spatial statistical processes but, as models become more realistic representations of a complex world, there is a need to express interactions between multiple processes. This is most directly seen by modeling vector-valued ``\nameref{s:geostat},'' $\{\mathbf{Y} (\mathbf{s}) : \mathbf{s} \in D^G\}$, and vector-valued ``\nameref{s:lattice},'' $\{\mathbf{Y} (\mathbf{s}_i) : {\mathbf{s}_i \in \DD^L}\}$, where the $k$-dimensional vector $\mathbf{Y (s)} \equiv (Y_1(\mathbf{s}), \dots, Y_k(\mathbf{s}))'$ represents the multiple processes at the generic location $\mathbf{s} \in \mathcal{D}$. Vector-valued spatial point processes, discussed in Section~\ref{s:point}, can be represented as a set of $k$ point processes, $\left\{ \{\mathbf{s}_{1,i}\},...,\{\mathbf{s}_{k,i}\} \right\}$, and these are presented in \citet[][Chapter 14]{Baddeley2015}. If we adopt a hierarchical-statistical-modeling approach, it is possible to construct multivariate spatial processes whose component univariate processes could come from any of the three types of spatial processes presented in the previous three subsections. This is because, at a deeper level of the hierarchy, a core {\em multivariate geostatistical process} can control the spatial dependance for processes of any type, which allows the possibility of hybrid multivariate spatial statistical processes.

In what follows, we  describe briefly two approaches to constructing multivariate geostatistical processes, one based on a joint approach and the other based on a conditional approach. We consider the case $k=2$, namely the bivariate spatial process $\{(Y_1\mathbf{(s)}, Y_2\mathbf{(s)})' :\mathbf{s}\in D^G \}$, for illustration. The joint approach involves directly constructing a valid spatial statistical model from $\boldsymbol\mu\mathbf{(s)} \equiv (\mu_1\mathbf{(s)}, \mu_2\mathbf{(s)})' \equiv (\EE(Y_1\mathbf{(s)}),\EE(Y_2\mathbf{(s)}))'$, for $\mathbf{s} \in \DD^G$, and from
\begin{equation}
\cov (Y_l\mathbf{(s)}, Y_m\mathbf{(u)}) \equiv \CC_{l m}\mathbf{(s,u)}; \;\;\;  l ,m =1,2 ,     \label{aa24}
\end{equation}
for $\mathbf{s}, \mathbf{u} \in \DD^G$.
The bivariate-process mean ${\boldsymbol\mu}(\cdot)$ is typically modeled as a vector linear regression; hence it is straightforward to model the bivariate mean once the appropriate regressors have been chosen.

Analogous to the univariate case, the set of covariance and cross-covariance functions, $\{\CC_{11}(\cdot,\cdot)$, $\CC_{22}(\cdot,\cdot)$, $\CC_{12}(\cdot,\cdot)$, $\CC_{21}(\cdot,\cdot)\}$, 
have to satisfy positive-definiteness conditions for the bivariate geostatistical model to be valid, and it is important to note that, in general, $\CC_{12}(\mathbf{s,u}) \neq \CC_{21}(\mathbf{s,u})$. There are classes of valid models that exhibit symmetric cross-dependance, namely $\CC_{12}(\mathbf{s,u}) = \CC_{21}(\mathbf{s,u})$, such as the linear model of co-regionalization \citep{gelfand2004nonstationary}. These are not  reasonable models for ore-reserve estimation when there has been preferential mineralization in the ore body.

The joint approach can be contrasted with a conditional approach \citep{cressie2016multivariate}, where each of the $k$  processes is a node of a directed acyclic graph that guides the conditional dependance of any process, given the remaining processes. Again consider the bivariate case (i.e., $k=2$), where there are only two nodes such that $Y_1(\cdot)$ is at node 1, $Y_2(\cdot)$ is at node 2, and a directed edge is declared from node 1 to node 2. Then the appropriate way to model the joint distribution is through 
\begin{equation}\label{a24a}
[Y_1(\cdot),Y_2(\cdot)]=[Y_2(\cdot)\mid Y_1(\cdot)][Y_1(\cdot)] ,
\end{equation}
where $[Y_2(\cdot)\mid Y_1(\cdot)]$ is shorthand for $[Y_2(\cdot)\mid \{ Y_1(\mathbf{s}) : \mathbf{s} \in D^G \}]$.

The geostatistical model for $[Y_1(\cdot)]$ is simply a univariate model based on a mean function $\mu_1(\cdot)$ and a valid covariance function $\CC_{11}(\cdot,\cdot)$, which was discussed in ``\nameref{s:geostat}.'' Now assume that $Y_2(\cdot)$ depends on $Y_1(\cdot)$ as follows: For $\mathbf{s,u} \in D^G$,
\begin{eqnarray}
\EE[Y_2(\mathbf{s})\mid Y_1(\cdot)] &\equiv& \mu_2(\mathbf{s}) + \int_{\DD^G} b(\mathbf{s,v}) \left(Y_1(\mathbf{v}) - \mu_1(\mathbf{v}) \right) \;\dd\mathbf{v} , \label{bb24a} \\  
\cov(Y_2(\mathbf{s}),Y_2(\mathbf{u})\mid Y_1(\cdot)) &\equiv& \CC_{2|1} (\mathbf{s,u}),        \label{bb24}
\end{eqnarray}
where $\CC_{2|1}(\cdot,\cdot)$ is a valid univariate covariance function and $b(\cdot,\cdot)$ is an integrable interaction function. The conditional-moment assumptions given by \eqref{bb24a} and \eqref{bb24} follow if one assumes that $\left( Y_1(\cdot), Y_2(\cdot) \right)'$ is a bivariate Gaussian process.

 \citet{cressie2016multivariate} show that, from \eqref{bb24a} and \eqref{bb24},
\begin{eqnarray}
\CC_{12}(\mathbf{s,u}) &=& \int_{D^G} \CC_{11} (\mathbf{s,v})b (\mathbf{u,v}) \;\dd\mathbf{v} \label{eq32}\\
\CC_{21}(\mathbf{s,u}) &=& \int_{D^G} \CC_{11} (\mathbf{v,u})b (\mathbf{v,s}) \;\dd\mathbf{v}  \\
\CC_{22}(\mathbf{s,u}) &=& \CC_{2\mid1}(\mathbf{s, u}) + \int_{D^G} \int_{D^G} b(\mathbf{s,v})\CC_{11}(\mathbf{v,w})b(\mathbf{u,w}) \;\dd\mathbf{v}\;\dd\mathbf{w} \label{eq34},
\end{eqnarray}
for $\mathbf{s,u} \in D^G$. Along with $\mu_1(\cdot)$, $\mu_2(\cdot)$, and $\CC_{11}(\cdot,\cdot)$, these functions \eqref{eq32}--\eqref{eq34} define a valid bivariate geostatistical process $[Y_1(\cdot),Y_2(\cdot)]$. A notable property of the conditional approach is that asymmetric cross-dependance (i.e., $\CC_{12}(\mathbf{s,u}) \neq \CC_{21}(\mathbf{s,u})$) occurs if $b(\mathbf{s,u}) \neq b(\mathbf{u,s})$. 

In summary, the conditional approach allows multivariate modeling to be carried out validly by simply specifying $\boldsymbol{\mu}(\cdot)= (\mu_1(\cdot),\mu_2(\cdot))'$ and two valid univariate covariance functions, $\CC_1(\cdot,\cdot)$ and $\CC_{2|1}(\cdot,\cdot)$. The strengths of the conditional approach are that only univariate covariance functions need to be specified \citep[for which there is a very large body of research; e.g.,][Chapter 4]{Cressie2011}, and that only integrability of $b(\cdot,\cdot)$,  the interaction function, needs to be assumed \citep{cressie2016multivariate}.

\begin{figure}
    \centering
    \includegraphics[scale = 1]{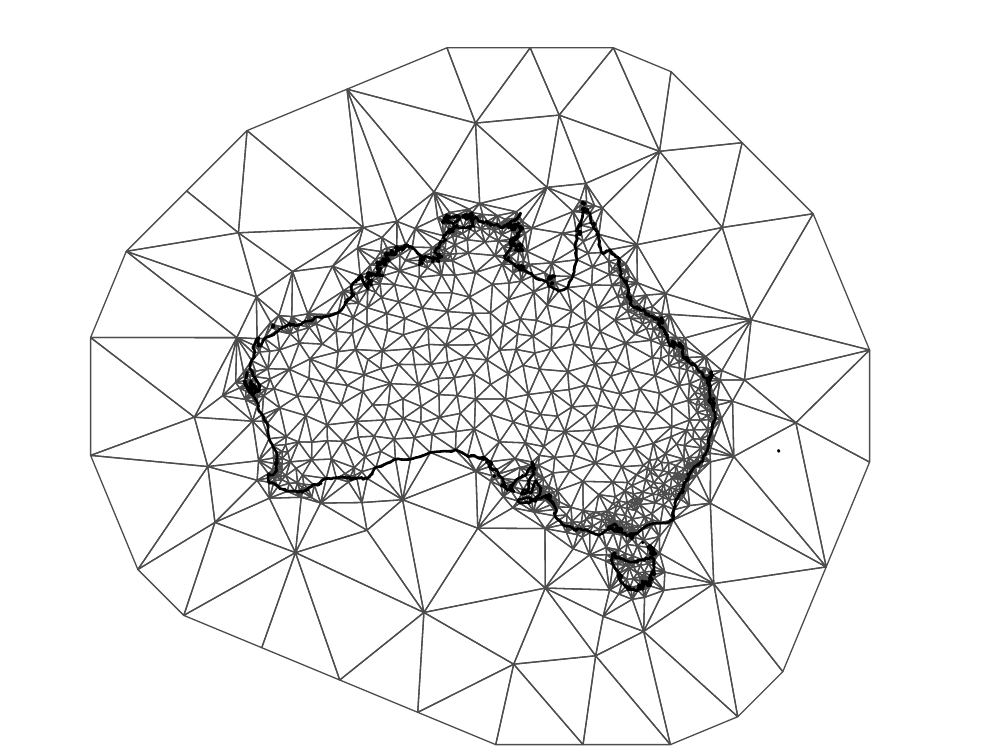}
    \caption{Discretization of the spatial domain $\mathcal{D}$, a convex region around Australia, into a triangular lattice.}\label{f:TriangularMesh}  
\end{figure}
\section{Spatial Discretization}\label{s:discrete}

Although geostatistical processes are defined on a continuous spatial domain $D^G$, this can limit the practical feasibility of statistical inferences due to computational and mathematical considerations. For example, kriging from an $n$-dimensional vector of data involves the inversion of an $n \times n$ covariance matrix, which requires  order  $n^3$ floating-point operations and  order  $n^2$  storage in available memory. These costs can be prohibitive for large spatial datasets; hence, spatial discretization to achieve scalable computation for spatial models is an active area of research.

In practical applications, spatial statistical inference is required up to a finite spatial resolution. Many approaches take advantage of this by dividing the spatial domain $\mathcal{D}$ into a lattice of discrete points in $\mathcal{D}$, as shown in Figure~\ref{f:TriangularMesh}. As a consequence of this discretization, a geostatistical process can be approximated by a lattice process, such as a Gaussian MRF \citep[e.g.,][Section 5.1]{Rue2005}, however sometimes this can result in undesirable discretization errors and artifacts. More sophisticated approaches have been developed to obtain highly accurate approximations of a geostatistical (i.e., continuously indexed) spatial process evaluated over an irregular lattice, as we now discuss. 

Let the original domain $\mathcal{D}$ be bounded and suppose it is tesselated into the areas $\{ A_j \subset \mathcal{D} : j = 1,\dots,m \}$ that are small, non-overlapping {\em basic areal units} \citep[BAUs;][]{Nguyen2012}, so that $\mathcal{D} = \cup_{j=1}^m A_j$, and $A_j \cap  A_k$ is an empty set for any $j \neq k \in \{1,\dots,m\}$; Figure~\ref{f:TriangularMesh} gives an example of triangular BAUs. Spatial basis functions $\{ \phi_\ell(\cdot):\; \ell = 1,\dots,r\}$, can then be defined on the BAUs. For example, \citet{Lindgren2011} used triangular basis functions where $r > m$, while fixed rank kriging \citep[FRK;][]{Cressie2008,ZammitMangion2020} can employ a variety of different basis functions for $r < m$, including multi-resolution wavelets and bisquares.

Vecchia approximations \citep[e.g.,][]{datta2016hierarchical,Katzfuss2020} are also defined using a lattice of discrete points $D^L \ \subset \ D^G \subset \mathcal{D}$, that include the coordinates of the observed data   $\DD^{G*} = \{\mathbf{s}_1, \dots, \mathbf{s}_n\}$ and the prediction locations $\{\mathbf{s}_{n+1}, \dots, \mathbf{s}_{n+p}\}$. Let $[ \mathbf{X}] \equiv [ \mathbf{Z}, Y]$, where data $\mathbf{Z}$ and spatial process $Y$ are associated with the lattice $D^L \equiv \{ \mathbf{s}_1, \dots, \mathbf{s}_n, \mathbf{s}_{n+1}, \dots, \mathbf{s}_{n+p} \}$. It is a property of joint and conditional distributions that this can be factorized into a product:
\begin{equation}\label{eq:jointCond}
[ \mathbf{X}] = \prod_{i=1}^{n+p} [X(\mathbf{s}_i) \mid X(\mathbf{s}_1), \dots, X(\mathbf{s}_{i-1}) ] .
\end{equation}
In the previous section, the set of spatial coordinates ${\DD}^L$ had no fixed ordering. However, a Vecchia approximation requires that an artificial ordering is imposed on $\{ \mathbf{s}_1, \dots, \mathbf{s}_{n+p} \}$. Let the ordering be denoted by $\{ \mathbf{s}_{(1)}, \dots, \mathbf{s}_{(n+p)} \}$, and define the set of neighbors $\mathcal{N}(\mathbf{s}_{(i)}) \subset \{\mathbf{s}_{(1)}, \dots, \mathbf{s}_{(i-1)}\}$, similarly to ``\nameref{s:lattice},'' except that these neighborhood relations are not reciprocal: If for $j < i$, $\mathbf{s}_{(j)}$ belongs to $\mathcal{N}(\mathbf{s}_{(i)})$, then $\mathbf{s}_{(i)}$ cannot belong to $\mathcal{N}(\mathbf{s}_{(j)})$. As part of the Vecchia approximation, a fixed upper bound $q \ll n$ on the number of neighbours is chosen. That is, $|\mathcal{N}(\mathbf{s}_{(i)})| \le q$, so that the lattice formed by $\{ \mathcal{N}(\mathbf{s}_i): i = 1,..., n + p \}$ is a directed acyclic graph, which results in a partial order in $\mathcal{D}$ \citep{cressie1998image}.

The joint distribution $[\mathbf{X}]$ given by \eqref{eq:jointCond} is then approximated by:
\begin{equation}\label{eq:Vecchia}
 \prod_{i=1}^{n+p} [ X(\mathbf{s}_{(i)}) \mid \mathbf{X}(\mathcal{N}(\mathbf{s}_{(i)}))] \equiv [ \widetilde{\mathbf{X}} ] ,
\end{equation}
which is a Partially-Ordered Markov Model \citep[POMM;][]{cressie1998image}. This Vecchia approximation, $[\widetilde{\mathbf{X}}] $, is a distribution coming from a valid spatial process on the original, uncountable, unordered index set $D^G$ \citep{datta2016hierarchical}, which means that it can be used as a geostatistical process model with considerable computational advantages. For example, it can be used as a random log-intensity function, $\log (\lambda(\mathbf{s}))$, in a hierarchical point-process model, or it can be combined with other processes to define models described in ``\nameref{s:multivariate}.'' However, in all of these contexts it should be remembered that the resulting predictive process, $[ \widetilde{\mathbf{X}} \mid \mathbf{Z} ]$, is an approximation to the true predictive process, $[ {\mathbf{X}} \mid \mathbf{Z} ]$.

\section{Spatio-Temporal Processes}\label{s:spacetime}
The section titled ``\nameref{s:multivariate}'' introduced processes that were written in vector form as,
\begin{equation}
 \mathbf{Y}\mathbf({s})\equiv(Y_1(\mathbf{s}),...,Y_k (\mathbf{s}))'; \; \mathbf{s} \in D^G.     \label{aaa4}
\end{equation}
In that section, we distinguished between the joint approach and the conditional approach to multivariate-spatial-statistical modeling and, under the conditional approach, we used a  directed acyclic graph to give a blueprint for the multivariate spatial dependance.

Now, consider a spatio-temporal process, 
\begin{equation}
\{Y(\mathbf{s};t) : \mathbf{s} \in D^G;  \  t\in \mathcal{T}\} ,   \label{bbb4}
\end{equation}
where $\mathcal{T}$ is a temporal index set. Clearly, if $\mathcal{T} = \{1,2,...\}$ then \eqref{bbb4} becomes a spatial process of time series, $\{Y(\mathbf{s};1), Y(\mathbf{s};2), \dots : \mathbf{s} \in D^G\}$. If $\mathcal{T} = \{1,2,\dots,k\}$, and we define $Y_j (\mathbf{s}) \equiv Y(\mathbf{s};j)$, for $j = 1, ...,k$, then the resulting spatio-temporal process can be represented as a multivariate spatial process given by \eqref{aaa4}. Not surprisingly, the same dichotomy of approach to modeling statistical dependance (i.e., joint versus conditional) occurs for spatio-temporal processes as it does for multivariate spatial processes.

Describing all possible covariances between $Y$ at any spatio-temporal ``location'' $(\mathbf{s};t)$ and any other one $(\mathbf{u};v)$, amounts to treating ``time'' as simply another dimension to be added to the d-dimensional Euclidean space, $\mathbb{R}^d$. Taking this approach, spatio-temporal statistical dependence can be expressed in $(d+1)$-dimensional space through the covariance function, 
\begin{equation}
C (\mathbf{s};t,\mathbf{u};v) \equiv \cov(Y(\mathbf{s};t), Y(\mathbf{u};v)); \qquad \mathbf{s}, \mathbf{u} \in D^G, \ t,v \in \mathcal{T} .        \label{ccc4}
\end{equation}
Of course, the time dimension has different units than the spatial dimensions, and its interpretation is different since the future is unobserved. Hence, the joint modeling of space and time based on \eqref{ccc4} must be done with care to account for the special nature of the time dimension in this \textit{descriptive approach} to spatio-temporal modeling.

From current and past spatio-temporal data $\mathbf{Z}$, predicting past values of $Y$ is called \textit{smoothing}, predicting unobserved values of the current $Y$ is called \textit{filtering}, and predicting future values of $Y$ is called \textit{forecasting}. The Kalman filter \citep{Kalman1960new} was developed to provide fast predictions of the current state using a methodology that recognises the ordering of the time dimension. Today's filtered values become ``old'' the next day when a new set of current data are received.  Using a dynamical approach that we shall now describe, the Kalman filter updates yesterday's optimal filtered value with today's data very rapidly, to obtain a current optimal filtered value.

The best way to describe the dynamical approach is to discretize the spatial domain. The previous section, ``\nameref{s:discrete},'' describes a number of ways this can be done; here we shall consider the discretization that is most natural for storing the attribute and location information in computer memory, namely a fine-resolution lattice $D^L$ of pixels or voxels (short for ``volume elements''). Replace $\{Y(\mathbf{s};t): \mathbf{s} \in D^G, \; t=1,2,\dots \}$ with $\{Y(\mathbf{s};t): \mathbf{s} \in D^L, \; t=1,2,\dots \}$, where $D^L\equiv\{\mathbf{s}_1,...,\mathbf{s}_m\}$ are the centroids of elements of small area (or small volume) that make up $D^G$. Often the areas of these elements are specified to be equal, having been defined by a regular grid. As we explain below, this allows a {\em dynamical approach} to constructing a statistical model for the spatio-temporal process $Y$ on the discretized space-time cube, $\{\mathbf{s}_1,..., \mathbf{s}_m\} \times \{1,2,...\}$. 

Define $\mathbf{Y}_t \equiv (Y(\mathbf{s};t):\mathbf{s} \in D^L)'$, which is an m-dimensional vector. Because of the temporal ordering, we can write the joint distribution of $\{Y (\mathbf{s};t):\mathbf{s} \in D^L, \; t=1,...,k\} $ from $t=1$ up to the present time $t=k$, as
\begin{equation}\label{dd4d}
[ \mathbf{Y}_1, \mathbf{Y}_2,..., \mathbf{Y}_k] = [\mathbf{Y}_1][\mathbf{Y}_2\mid \mathbf{Y}_1] ... [ \mathbf{Y}_k\mid\mathbf{Y}_{k-1},..., \mathbf{Y}_2, \mathbf{Y}_1 ],
\end{equation}
which has the same form as \eqref{eq:jointCond}. Note that this conditional modeling of space and time is a natural approach, since time is completely ordered. The next step is to make a Markov assumption, and hence \eqref{dd4d} can be written as
\begin{equation}
[ \mathbf{Y}_1, \mathbf{Y}_2,..., \mathbf{Y}_k] = [\mathbf{Y}_1] \prod_{j=2}^{k} [\mathbf{Y}_j\mid\mathbf{Y}_{j-1} ].        \label{eee4}
\end{equation}
This is the same Markov property that we previously discussed in ``\nameref{s:lattice},'' except it is now applied to the completely ordered one-dimensional domain, $\mathcal{T} = \{1,2,\dots\}$, and $\mathcal{N}(j) = j-1$. The Markov assumption makes our approach dynamical: It says that the present, conditional on the past, in fact only depends on the ``most recent past.'' That is, since $\mathcal{N}(j) = j-1$, the factor $[\mathbf{Y}_j \mid \mathbf{Y}_{j-1},...,\mathbf{Y}_2,\mathbf{Y}_1] = [\mathbf{Y}_j \mid \mathbf{Y}_{j-1}]$, which results in the model \eqref{eee4}.

For further information on the types of models used in the descriptive approach given by \eqref{ccc4} and the types of models used in the dynamical approach given by \eqref{eee4}, see \citet[][Chapters 6--8]{Cressie2011} and \citet[][Chapters 4 and 5]{Wikle2019}. The statistical analysis of observations from these processes is known as {\em spatio-temporal statistics}. Inference (estimation and prediction) from spatio-temporal data using R software can be found in \cite{Wikle2019}.                                              

\section{Conclusion}\label{s:trends}

Spatial-statistical methods distinguish themselves from spatial-analysis methods found in the geographical and environmental sciences, by providing well calibrated quantification of the uncertainty involved with estimation or prediction. Uncertainty in the scientific phenomenon of interest is represented by a spatial {\em process model}, $\{ Y(\mathbf{s}) : \mathbf{s} \in \DD \}$, defined on possibly random $\DD$ in $\mathbb{R}^d$, while measurement uncertainty in the observations $\mathbf{Z}$ is represented in a {\em data model}. In ``\nameref{s:intro},'' we saw how these two models are combined using Bayes' Rule \eqref{8a}, or the simpler version \eqref{9a}, to calculate the overall uncertainty needed for statistical inference. 

With some exceptions \citep[e.g.,][]{cressie2003}, spatial-statistical models \eqref{1aaa} rarely consider the case of measurement error in the locations in $\mathcal{D}$. Here we focus on a spatial-statistical model for the location error: Write the observed locations as $D^* \equiv\{ \mathbf{u}_i: i = 1,...,n \}$; in this case, a part of the data model is $[D^* \mid D]$, and a part of the process model is $[\DD]$. Finally then, the data
consist of both locations and attributes and
 are $\mathbf{Z} \equiv \{ (\mathbf{u}_i, Z(\mathbf{u}_i)): i = 1,\dots,n \}$, the spatial process model is $[ Y, D ]$, and the data model is $[ \mathbf{Z} \mid Y, D ]$. Then Bayes' Rule given by \eqref{8a} is used to infer the unknown $Y$ and $D$ from the predictive distribution $[Y,D \mid \mathbf{Z}]$.

There are three main types of spatial process models: geostatistical processes where uncertainty is in the process $Y$, which is indexed continuously in $D = D^G$; lattice processes where uncertainty is also in $Y$, but now $Y$ is indexed over a countable number of spatial locations $D = D^L$; and point processes where uncertainty is in the spatial locations $D = D^P$. Multiple spatial processes can interact with each other to form a multivariate spatial process. Importantly, processes can vary over time as well as spatially, forming a spatio-temporal process. 

As the size of spatial datasets have been increasing dramatically, more and more attention has been devoted to scalable computation for spatial-statistical models. Of particular interest are methods that use ``\nameref{s:discrete}'' to approximate a continuous spatial domain, $D^G$. There are other recent advances in spatial statistics that we feel are important to mention, but their discussion here is necessarily brief. 

Physical barriers can sometimes interrupt the statistical association between locations in close spatial proximity. Barrier models \citep{bakka2019non} have been developed to account for these kinds of discontinuities in the spatial correlation function. Other methods for modeling nonstationarity, anisotropy, and heteroskedasticity in spatial process models are an active area of research. 
                                                                                                                                                                        
It can often be difficult to select appropriate prior distributions for the parameters of a stationary spatial process, for example its correlation-length scale. Penalised complexity (PC) priors \citep{simpson2017penalising} are a way to encourage parsimony by favoring parameter values that result in the simplest model consistent with the data. The likelihood function of a point process or of a non-Gaussian lattice model can be both  analytically and computationally intractable. Surrogate models, emulators, and quasi-likelihoods have been developed to approximate these intractable likelihoods \citep{moores2020bayesian}. 

Copulas are an alternative method for modeling spatial dependence in multivariate data, particularly when the data are non-Gaussian \citep{krupskii2019copula}. One area where non-Gaussianity can arise is in modeling the spatial association between extreme events, such as for temperature or precipitation \citep{tawn2018modelling,Bacro2020}.

As a final comment, we reflect on how the field of geostatistics has evolved, beginning with applications of spatial stochastic processes to mining: In the 1970s, Georges Matheron and his Centre of Mathematical Morphology in Fontainebleau were part of the Paris School of Mines, a celebrated French tertiary-education and research institution. To see what the geostatistical methodology of the time was like, the interested reader could consult \citet{Jou1978}, for example. Over the following decade, geostatistics became notationally and methodologically integrated into statistical science and the growing field of spatial statistics \citep[e.g.,][]{Ripley1981,Cressie1993}. It took one or two more decades before geostatistics became integrated into the hierarchical-statistical-modeling approach to spatial statistics \citep[e.g.,][Chapter 4]{Cressie2011}. The presentation given in our review takes this latter viewpoint and explains well known geostatistical quantities such as the variogram and kriging in terms of this advanced, modern view of geostatistics. We also include a discussion of uncertainty in the spatial index set as part of our review, which offers new insights into spatial-statistical modeling. Probabilistic difficulties with geostatistics, of making inference on a possibly non-countable number of spatial random variables from a finite number of observations, can be finessed by discretizing the process. In a modern computing environment, this is key to doing spatial-statistical inference (including kriging).

\vspace{10pt}
\noindent {\bf Acknowledgments}: Cressie's research was supported by an Australian Research Council Discovery Project (Project number DP190100180). Our thanks go to Karin Karr and Laura Cartwright for their assistance in typesetting the manuscript.


\bibliography{refs}

\end{document}